# Photoactivation of color centers induced by laser irradiation in ion-implanted diamond

V. Pugliese[1], E. Nieto Hernández[1,#], E. Corte[1], M. Govoni[2], S. Ditalia Tchernij[1], P. Olivero[1], J. Forneris[1]

[1] *Physics Department, University of Torino, 10125 Torino Italy, and Istituto Nazionale di Fisica Nucleare (INFN), Sezione di Torino, 10125 Torino, Italy*
[2] *Department of Physics, Computer Science and Mathematics, University of Modena and Reggio Emilia, 41125 Modena Italy*

[#] corresponding author: elena.nietohernandez@unito.it

**Abstract**

Split-vacancy color centers in diamond are promising solid state platforms for the implementation of photonic quantum technologies. These luminescent defects are commonly fabricated upon low energy ion implantation and subsequent thermal annealing. Their technological uptake will require the availability of reliable methods for the controlled, large scale production of localized individual photon emitters. This task is partially achieved by controlled ion implantation to introduce selected impurities in the host material, and requires the development of challenging beam focusing or collimation procedures coupled with single-ion detection techniques. We report on protocol for the direct optical activation of split-vacancy color centers in diamond via localized processing with continuous wave laser at mW optical powers. We demonstrate the activation of photoluminescent Mg– and Sn-related centers at both the ensemble and single-photon emitter level in ion-implanted, high-purity diamond crystals without further thermal processing. The proposed lithographic method enables the activation of individual color centers at specific positions of a large area sample by means of a relatively inexpensive equipment offering the real-time, *in situ* monitoring of the process.

1.  **Introduction**

Color centers in diamond represent promising platforms for the implementation of quantum technologies [1], [2], [3]. Their capability to generate single photons on-demand represents a viable tool for realizing high-density photonic platforms operating at room temperature [4], [5]. Among the known single-photon emitters (SPEs) in diamond, the negatively charged nitrogen-vacancy (NV$^-$) center [6], [7], [8], offers enticing applications in the field of quantum sensing and metrology in virtue of its outstanding optically addressable spin properties at room temperature [9], [10] and its sensitivity to external electromagnetic fields [11], [12]. The latter feature, combined with a low Debye-Waller factor and a long radiative lifetime, could nevertheless represent a significant drawback for other specific applications in the framework of quantum technologies, in which the generation of indistinguishable photons at a high rate is required [1], [13]. Various alternative color centers have therefore drawn increasing attention in the last few years, including the group-IV related defects (also known as G4V color centers) [14], [15], [16], [17], [18] and the magnesium-vacancy emitter [19]. These systems, all of which are based on the split-vacancy structural configuration, display substantially better properties in terms of zero-phonon line (ZPL) emission linewidth, Debye-Waller factor and emission rate, and they offer lower environmental sensitivity in addition to energy level configurations that allow the implementation of coherent control schemes [4], [20].

In particular, this work focuses on the formation of two of the afore-mentioned luminescent defects in diamond, namely the negative charge state of the tin-vacancy (SnV) [21], [22] and magnesium-vacancy (MgV) defect complexes [19]. The SnV$^-$ center has emerged as an important quantum platform in the field of quantum communication due to its long coherence time, even at relatively high temperatures [15], [23]. On the other hand, the MgV$^-$ center is a newly discovered emitter that holds substantial potential for implementation in quantum sensing and photonics [19], [24]. Understanding the dynamics related to the formation and activation of these defects is therefore of crucial importance for fully exploiting their quantum-opto-physical properties in quantum technologies.

To date, the most widely employed method for creating defects in diamond is based on ion implantation followed by a high-temperature annealing to promote the formation of stable defects in the desired structural configuration [25], [26], [27], [28]. This procedure enables a precise control of the location and quantity of impurities introduced into the diamond substrate [29], [30], [31]. However, the defect formation efficiency (defined as the ratio between the areal density of optically-active color centers generated by ion implantation and ion fluence in an undoped substrate) after conventional thermal annealing has been reported to be critically low, i.e. not exceeding 5% and 8% for SnV$^-$ and MgV$^-$ optically active charge states, respectively [32].

Interestingly, recent emission channeling experiments have determined that the structural formation efficiency, i.e. the fraction of implanted ions resulting in a bond-centered configuration in the diamond lattice, exceeds 30% for both Sn- [33] and Mg-containing defects [19] in as-implanted samples, i.e. before any thermal processing stage. This observation is compatible with the split-vacancy configuration of the MgV and SnV defect complexes, although at the state of the art it is not fully understood how probable is the formation of the defect in such configuration with respect to other possible defect arrangements based on the incorporation of one (or more) lattice vacancies. Nevertheless, the formation of split-vacancy centers in a far-from-negligible concentration can be hypothesized. The fact that such emitters are largely inactive in as-implanted samples suggests that the formation of photon emitting color centers requires a further processing step, functional to the stabilization of the electronic configuration of the defect in the desired optically-active state.

This work explores the optical activation of MgV$^-$ and Sn-related (PL emission at 595 nm in Sn-implanted diamond [34], [21]) optically active color centers in diamond straight after ion implantation ("as-implanted", in the following), i.e. without any post-implantation thermal treatment. The protocol discussed here relies on the localized activation of color centers using a continuous-wave laser with mW power. This simple and accessible process enables stable *in-situ* activation of color centers through exposure to a conventional and widely available instrumentation. This arrangement is not only convenient due to the use of the same setup used for optical characterization, but also provides insights into the mechanisms that determine the optical activation of color centers and their formation.

**Results**

**Laser processing.** The experiments were performed on two high-purity diamond samples, 2 x 2 x 0.5 mm$^3$ sized, IIa-type, single-crystal diamond plates produced by ElementSix by chemical vapor deposition synthesis. The first sample (referred to as "Sample A" in the following) was implanted with MgH$^-$ ions at an energy of 80 keV, and an ion fluence of 1x10$^{13}$ cm$^{-2}$. The second sample ("Sample B" in the following) was implanted with Sn$^-$ ions at an energy of 56 keV and an fluence of 1x10$^{12}$ cm$^2$. The surfaces of both samples were untreated, i.e. no specific surface termination was induced by means of chemical or plasma processes, either preliminarily or subsequently to the ion implantation process.

The as-implanted samples were characterized in a custom fiber-coupled confocal microscope. A 522 nm diode laser (referred as "probing laser" in the following) with a fixed optical power of 100 µW was focused on the sample surface in the photoluminescence (PL) mapping and spectral analysis. The above-reported optical power value was defined as low enough to avoid giving rise to any photoactivation effects. The photoactivation of MgV$^-$ and Sn-related emission in as-implanted samples was studied as a function of the wavelength of the laser beams adopted for the substrate processing. To this purpose, three continuous wave (CW) lasers (referred as "processing" lasers) emitting at 405 nm, 445 nm, 522 nm wavelengths were employed in a "high" (i.e. 0.1-25 mW) power range. Processing and probing lasers were combined with a long-pass dichroic mirror (520 nm cutoff wavelength), allowing a simultaneous irradiation of the substrates.

The investigation involved the analysis of the dependence of photoactivation on the optical power and the duration of the laser exposure for each considered activation wavelength. To this aim, a systematic study of the implanted samples was performed by defining, for each processing-laser wavelength, a regular grid of spots processed with different laser irradiation parameters at varying optical powers (i.e. 8 values in the 0.1-25 mW range) and exposure times (i.e. 6 values ranging from 1 to 75 minutes). This arrangement allowed to assess the dependence of the photoactivation efficiency as a function of the energy delivered to the sample via the processing laser irradiation. Additionally, a longer irradiation (10 hours) was performed at the maximum available power for each activation wavelength to identify the asymptotic behavior of the photoactivation process.

For each of the irradiated grids a PL map was acquired before and after the processing-laser irradiation. A spectral analysis was carried out for each irradiated spot to identify the origin of the PL emission. The spectral range considered for the study of the MgV$^-$ and the Sn-related defects was 550-650 nm and 580-680 nm, respectively. In both cases, the spectra were processed by a background PL subtraction with respect to the PL collected from an unirradiated area of the samples, and were normalized to the first-order Raman peak intensity for sake of consistency, in consideration of possible slight variations in probing-laser focussing conditions.

**Effects of exposure time and laser irradiation power.** Two exemplary results are reported in **Figures 1 and 2** for the photoactivation of MgV$^-$ and Sn-related centers, respectively, under 405 nm processing-laser irradiation. **Figs. 1a** and **2a** show the PL maps acquired in implanted diamond following the patterning of a grid of regularly spaced spots under different optical powers (increasing across rows) and exposure time (increasing across columns). The processed spots are associated with an increase in the PL emission intensity with respect to the background, unequivocally indicating the local activation of color centers. A systematic analysis of the spectral emission from each activated spot (**Figs. 1b** and **2b**) evidenced the signature of specific color centers, namely the MgV$^-$ center in Sample A, denoted by a sharp ZPL at 557.4 nm, and the 595 nm Sn-related peak in Sample B [21], [34]. By contrast, the background emission acquired from the implanted, untreated region surrounding the lased spots (orange lines in **Figs. 1b** and **2b**) evidenced the lack of those signatures and the sole presence of the first-order Raman peak at 1332 cm$^{-1}$ wavenumber shift.

The PL emission rate was found to increase in both samples with the optical power of the processing laser, indicating a clear correlation between the sample exposure and the activation yield of color centers. This correlation relation can be appreciated in **Figs. 1c** and **2c**, where the area underlying the emission spectra profiles in the 550-650 nm and 580-680 nm range for the

MgV⁻ and Sn-related emission is plotted against the 405 nm processing optical power for the two limit exposure times (1 min and 75 min). Particularly, the data exhibit a clearly increasing trend, which is accompanied by a saturation behavior at high optical powers (i.e. > 10 mW for the MgV⁻ center and >20 mW for the Sn-related center).

Along with the power dependence study, the dependence of the PL emission intensity against the processing-laser exposure time was investigated for each of the considered optical powers. **Figs. 1d** and **2d** show the intensity of each activated spot (measured as the integral of the PL spectral peaks) for MgV⁻ and Sn-related emitters; integrative data are reported for different time exposures in the exemplary case of 405 nm processing laser wavelength under two different optical powers. In both cases an increasing trend with the exposure time is observed, with the MgV⁻ and Sn-related emission reaching a saturation value after 30 min and 50 min of processing-laser irradiation, respectively.

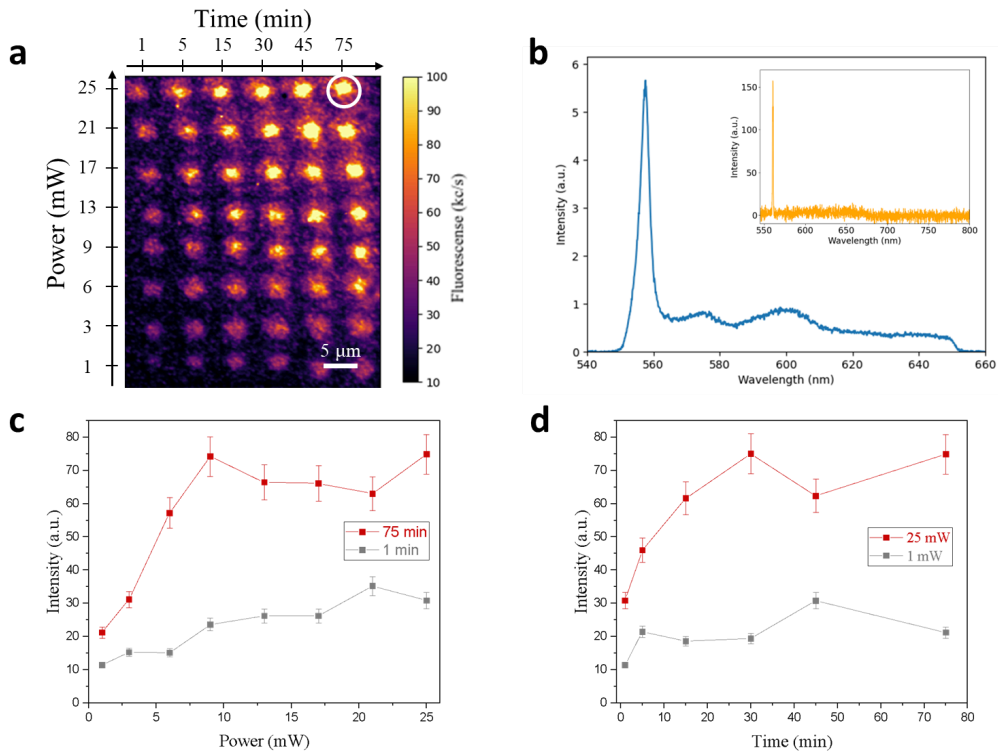

**Figure 1: a)** PL map of the array of spots activated in MgH⁻-implanted diamond upon 405 nm laser processing at varying conditions of optical power and exposure time. **b)** Emission spectrum, background subtracted, and normalized to first-order Raman peak, of the spot circled in white in (a) processed for 75 min at a 25 mW optical power. The inset (orange) displays the PL spectrum acquired from a region of sample implanted with MgH⁻ ions which did not undergo laser processing. **c)** Emission intensity of the MgV⁻ ZPL as a function of the laser processing optical power for exposure times of 1 min (grey) and 75 min (red). **d)** Emission intensity of the MgV⁻ ZPL as a function of the laser processing time under 1 mW (grey) and 25 mW (red) optical powers.

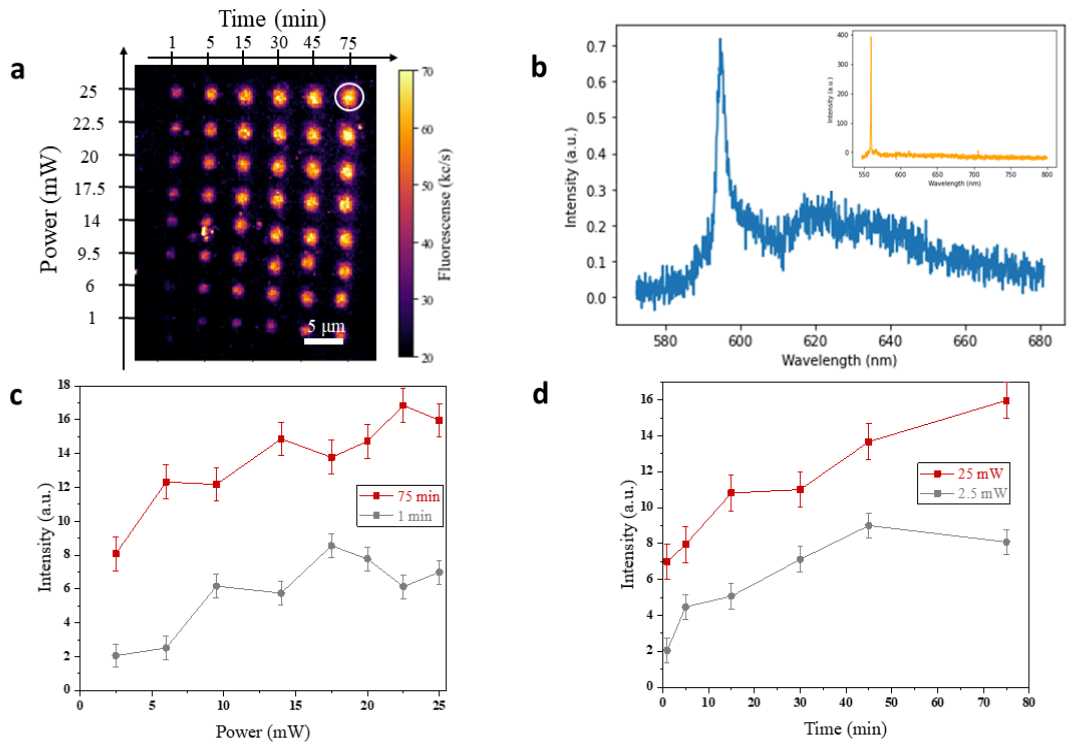

**Figure 2: a)** PL map of the array of spots activated in Sn⁻-implanted diamond upon 405 nm laser processing at varying conditions of optical power and exposure time. **b)** Emission spectrum, background subtracted, and normalized to first-order Raman peak, of the spot circled in white in (a) processed for 75 min at a 25 mW optical power. The inset (orange) displays the PL spectrum acquired from a region of sample implanted with Sn⁻ ions which did not undergo laser processing. **c)** Emission intensity of the 595 nm ZPL as a function of the laser processing optical power for exposure times of 1 min (grey) and 75 min (red). **d)** Emission intensity of the 595 nm ZPL as a function of the laser processing time under 2.5 mW (grey) and 25 mW (red) optical powers.

According to the emission channeling measurements available for Sn- and Mg-ion implanted samples [19], [33], we interpret the observed results as a process of laser-induced optical activation of ensembles of defects which are already available in the diamond lattice in the split-vacancy configuration, but that are configured in an electronic state which is associated with a dark state. This attribution is motivated by the fact that additional interpretations can be ruled out, as detailed in the following. Firstly, the considered wavelengths and instantaneous optical powers are insufficient to introduce any structural modification to the diamond crystal [35], or any thermal effect comparable with the typical annealing temperatures required for vacancy diffusion [36]. This latter consideration was further supported by the observation of the same activation process at cryogenic temperatures, particularly in consideration of the very large thermal conductivity of the diamond crystal. Moreover, the possible dependence of the photoactivation on surface functionalization was analyzed by performing laser processing treatments on surface oxidized reference samples, leading to no apparent changes in the trend of the PL emission increase with respect to the one observed for the untreated surface sample. This test enabled to rule out the possible attribution of the conversion process to a laser-assisted surface charge modification [37,38]. Moreover, it is worth stressing that this latter hypothesis was further in contrast with the experimental observation of the phenomenon upon laser-processing at cryogenic temperatures in vacuum conditions, at which laser-assisted surface chemistry is expected to be strongly inhibited.

**MgV⁻ photoactivation.** In this case, we can firstly assume that the photoactivation of MgV defect complexes in the optically active MgV⁻ charge state is due to the conversion from a single optically inactive configuration via photon absorption. Considering the lack of GR1 emission, i.e. the PL signature of the neutral vacancy, in the as-implanted sample (see **Figs. 1b** and **2b**), the radiation-induced vacancy density generated by ion implantation is then assumed to be configured in its negative charge state (ND1 center [39]), which is optically inactive in the considered spectral range (500-800 nm). This observation suggests that the Fermi level in the ion-implanted sample is located at minimum 3.2 eV above the valence band. In this case, the formation energy of the ND1 charge state of the vacancy is indeed more favorable with respect to the GR1 configuration [40]. This estimation of the Fermi level position in the energy gap is compatible with the doubly-negative charge state ($MgV^{2-}$) being the favored electronic configuration of the MgV defect complex, on the basis of the theoretical results obtained in Ref. [24]. Such interpretation is consistent with the lack of observed PL from the $MgV^-$ charge state in the as-implanted sample, and is reasonable considering the introduction of overall negative charge in the material via ion implantation. The increase in the $MgV^-$ emission upon laser processing is thus interpreted as the ionization of the $MgV^{2-}$ charge state via the transition of an electron to the conduction band.

The above-described process can be modeled as follows. The total density of Mg-containing split-vacancy defects $n_{SV}$ is given by the sum of the active and inactive configurations. Such configurations are respectively attributed to the sole $MgV^-$ and $MgV^{2-}$ charge states ($n_1$, $n_2$, respectively), i.e.:

$$n_1 + n_2 = n_{SV} \qquad (1)$$

The photoinduced charge state conversion between the "1-" and "2-" configurations can be thus described by the following time dependent rate equation:

$$\frac{dn_1(t)}{dt} = r_{21} n_2(t) - r_{12} n_1(t) \qquad (2)$$

where the $r_{21}$ ($r_{12}$) terms indicate the activation (deactivation) rate coefficients related to the conversion from (to) the $MgV^{2-}$ to (from) the $MgV^-$ configuration. By combining Eqs. (1-2) and implementing the experimental constraint $n_1(t=0)=0$, the expression for the density of MgV centers in the singly-negative charge state as a function of the exposure time to the processing laser is given by:

$$n_1(t) = \frac{r_{21}}{r_{21}+r_{12}} n_{SV} (1 - e^{-t(r_{21}+r_{12})}) \qquad (3)$$

Here, $r_{21}$ and $r_{12}$ can be regarded in general as transition rates dependent on the wavelength $\lambda$ and power P. The asymptotic behavior at t→∞ is indicative of the maximum fraction $r_{21}/(r_{21} + r_{12})$ of split-vacancy centers which can be converted in the $MgV^-$ charge state.

If the photoactivation is considered as a one-photon absorption process, the transition rates can be described in simple terms as [41] $r_{21} = a_{21}(\lambda) P$ and $r_{12} = a_{12}(\lambda) P$, leading to a power-independent fraction of emitters activated in the asymptotic limit (i.e. $n_1(t\to\infty) = a_{21}/(a_{21} + a_{12}) n_{SV} = A\, n_{SV}$). Conversely, at lower time scales, the product of the CW lasing power and the exposure time enables to describe the process in terms of the total energy $E$ deposited within the diffraction limited focal point of the optical objective:

$$n_1(E) = A\, n_{SV}(1-\exp(-(a_{21}+a_{12})\, E)) \qquad (4)$$

Under the assumption that the physical observable of the experiment - i.e. the PL emission intensity $I$ - is proportional to $n_1$, then the dependence of the latter from the energy delivered to the sample can be assessed. It is therefore worth noting that the functional form obtained in Eq. (4) offers a reasonable description of the data points reported in **Figure 1**. The wavelength dependence of the $I(E)$ photoactivation process was investigated for the MgV⁻ optically active charge state by generating an array of spots under different laser processing parameters, similarly to what presented in **Fig. 1a** for 405 nm CW irradiation. The processing was performed with CW optical powers in the mW range at exposure times varying from 1 min to 75 min, depending on the total energy deposited at each spot. A full list of the specific parameters is reported in **Table 1**. To prevent cross-effects with the other spots, the processes involving the highest energy deposition (corresponding to ~10 h laser exposure) were fabricated on separate regions of sample A, again with the purpose of assessing the asymptotic behavior of the photoactivation process at large deposited energies.

| Processing laser wavelength (nm) | Processing optical power range (mW) | Deposited energy range (J) |
|---|---|---|
| 405 | 1.0-25.0 | 0.1-900.0 |
| 445 | 0.8-5.0 | 0.1-136.8 |
| 522 | 1.5-7.7 | 0.1-151.2 |

**Table 1**. Laser processing parameters adopted for the photoactivation MgV⁻ and Sn-related centers.

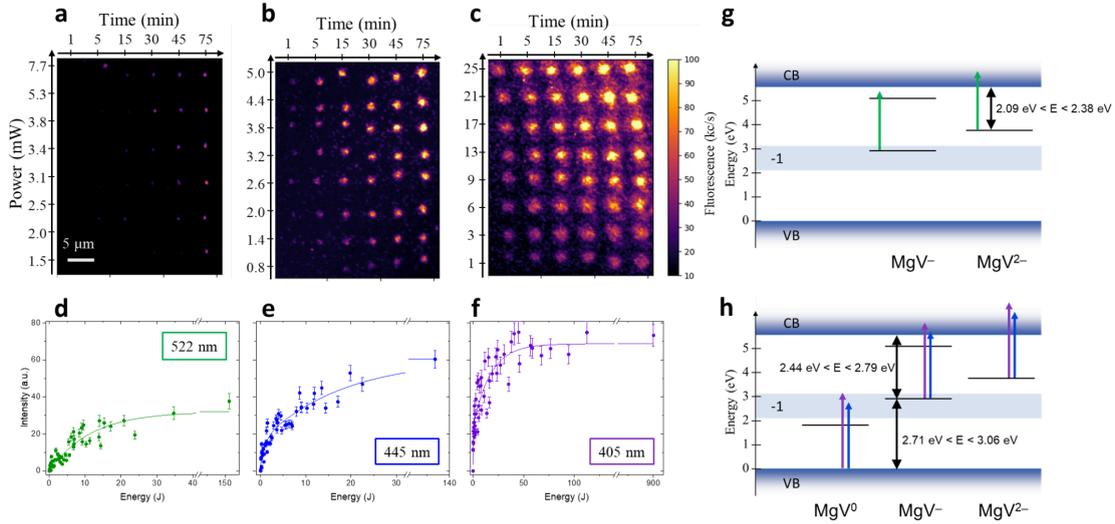

**Figure 3:** Photoactivation dependence on the laser processing parameters for the MgV⁻ color center. The first row shows the confocal maps acquired on an array of spots undergone photoactivation via laser processing at **a)** 522 nm, **b)** 445 nm, **c)** 405 nm. The second row shows the integrated PL intensity (integral of the MgV⁻ ZPL spectral peak) collected spots as a function of the energy deposited on each spot under processing wavelength of **d)** 522 nm,, **e)** 445 nm, **f)** 405 nm. **g)** Schematic representation of the energy levels of the MgV defect charge states. The negative charge state MgV⁻ is represented as a 2-level system in the diamond energy gap. Only the ground state of the doubly-negative charge state MgV²⁻ under the assumption that any excited state lies above the bottom of the conduction band. The colored green arrows represent the transitions induced by the processing laser at 522 nm. The shaded blue area in the gap depicts the Fermi Energy range of stability of the MgV⁻ configuration, according to [23]. **h)** Photo-assisted transitions induced by 405 nm and 445 nm laser processing, involving the neutral charge state of the MgV defect.

A confocal PL map acquired under 522 nm probing excitation is reported in **Figs. 3a-c** for each of the arrays fabricated under 522 nm, 445 nm and 405 nm. The PL intensity is encoded in the same color scale for all maps to highlight the effect of different processing wavelengths. The dependence of the PL intensity (measured as the integral of the MgV⁻ ZPL spectrum acquired from each processed spot) on the energy delivered to the spots is plotted in **Figs. 3d-f**, respectively. All the data points resulted to be distributed on the same curve independently of the specific optical power at which the processing laser was held, thus justifying the assumptions made in **Eq. (4)**. These graphs report a clear dependence of the asymptotic saturation plateau at large energies with the processing wavelength. Particularly, the saturation of the PL intensity is maximum under 405 nm processing and exhibits a decreasing trend at increasing wavelength, with a saturation intensity which is more than halved under 522 nm CW irradiation. Further processing performed with a 594 nm laser (not shown here) at similar optical powers did not result in the formation of any MgV$^-$ emitters. Therefore, it can be inferred that the photoactivation process is characterized by a threshold energy comprised between 2.09 eV and 2.38 eV, as schematically shown in **Fig. 3g**.

In order to validate the photophysical model proposed in Eqs. (1-4), the datasets in **Figs**. **3d-f** were fitted according to a single exponential function in the form $I(E)=I_0 (1-\exp(-E/\alpha))$. The fitting curve suitably accounted for the experimental data collected under 522 nm laser processing (**Fig. 3d**), but offered unsatisfactory results for those obtained at lower wavelengths. A reasonable agreement with the data was found for the 445 nm and 405 nm processing wavelengths when adding second exponential term to the fitting function, i.e.:

$I(E)=I_0 (1-a \exp(-E/\alpha)- b \exp(-E/\beta))$ (5)

The relevant parameter $I_0$, describing the fraction of emitters activated in the asymptotic limit, is listed in **Table 2** for each considered processing wavelength. This quantity, which is proportional to the ionization rate of the MgV$^{2-}$ charge state, shows a monotonic decreasing trend at increasing wavelengths, suggesting a more effective process at shorter wavelengths.

It is worth noting that the appearance of a second exponential term in the $I(E)$ curves at 445 nm and 405 nm is not justified in terms of non-linear photon absorption such as two-photon processes [41,42], particularly since the increase in the laser processing energy should favor one-photon ionization [42]. Conversely, here we assume the presence of an energy threshold (in the 2.38→2.79 eV range, i.e. between 522 nm and 445 nm) enabling the interaction of the MgV$^-$ configuration with a possible third charge state other than the MgV$^{2-}$. We can hypothesize that this level could be represented by the neutral charge configuration of the defect (MgV$^0$), although a definitive assessment would require further investigation. The ionization of the MgV$^-$ center via one-photon absorption in the $E_I$=2.38→2.79 eV range determines a complementary (i.e. Egap-EI=2.71→3.06 eV photon energy range, see **Fig. 3h** for a schematic representation) energy required for the reverse process that consists in the conversion of MgV$^0$ to MgV$^-$ via the capture of valence-band electrons. The occurrence of both ionization and recombination between MgV$^0$ and MgV$^-$ centers could explain thus the modest excitability of the MgV$^-$ emission under laser wavelengths below 500 nm reported in previous works [19]. It is finally worth remarking that the fitting value found for the $\alpha$ parameter with an energy scale of ~10 J is mutually compatible for all of the adopted processing wavelengths. Furthermore, the $\beta$ parameter identified for the 445 nm and 405 nm describes the additional laser-induced interactions as a significantly faster process described by a ~1 J energy scale.

| Wavelength (nm) | $I_0$ (a.u.) | $a$ (a.u.) | $b$ (a.u.) | $\alpha$ (J) | $\beta$ (J) |
|---|---|---|---|---|---|
| 520 | 32±4 | - | - | 12±3 | - |
| 445 | 61±9 | 44±9 | 15±5 | 19±8 | 1.2±0.6 |
| 405 | 66±4 | 46±4 | 13±6 | 18±5 | 0.4±0.4 |

**Table 2**. Fitting parameters of the wavelength-dependent $I(E)$ curves for the MgV$^-$ ZPL emission according to Eq. (5)

The time-dependent increase in the PL emission from Mg-implanted diamond was also investigated. **Fig. 4a** shows the PL intensity during the laser processing over a 900 s time interval under 522 nm wavelength at 5.3 mW optical power. The photon count rate was monitored using the same processing laser as PL excitation source in the sample. Furthermore, it is worth noting the difference due to the laser irradiation on the count traces acquired simultaneously with the photoactivation process. The count trace exhibits a monotonic increase in the PL emission intensity, indicating a cumulative activation of MgV⁻ centers located within the confocal excitation spot. Notably, the PL increase features several discrete jumps between constant photon count rates, suggesting the separate activation of individual color centers. The obtained activation resulted to be fully irreversible for all of the explored laser processing conditions, as confirmed by the juxtaposition (**Figs. 4b-c**) of two separate PL confocal maps acquired from the same region processed under 405 nm laser at a 4 months time interval. The persistence of the PL features, which is in line with the comparable formation energy of the three considered charge states within the stability range of the MgV⁻ configuration [23], guarantees a permanent activation of color centers at specific locations of the implanted samples, thus offering a viable tool for the direct writing (i.e. direct photo-activation) of single-photon emitters.

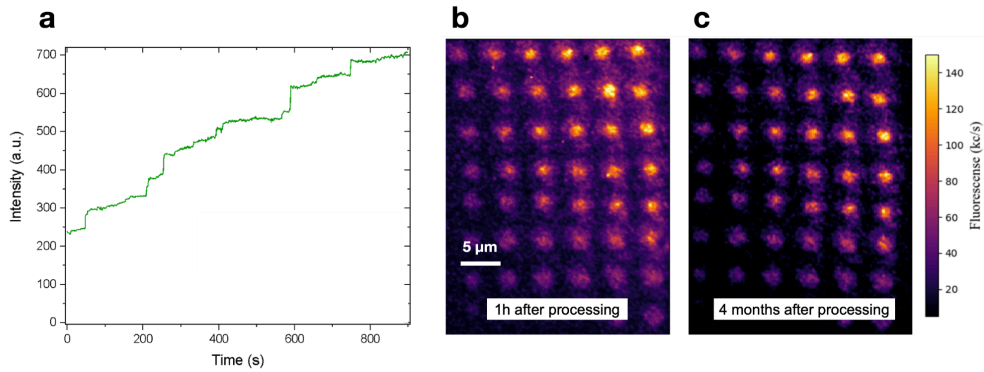

**Figure 4: a)** Real-time MgV⁻ PL intensity under laser processing at 522 nm wavelength (5.3 mW optical power). **b)-c)** Confocal PL microscopy scans of the region processed with 405 nm laser. The maps were acquired under 522 nm laser excitation (100 µW optical power) **b)** one hour and **c)** 4 months after the laser processing.

The observation of discrete jumps in the time-dependent PL emission during laser processing enabled to assess whether the CW treatment was effective at producing individual MgV⁻ optical centers. An exemplary case is displayed in Figure 8 in the proximity of a spot processed for 10 hours under the 445 nm laser wavelength (optical power: 5 mW). The corresponding PL map (**Fig. 5a**, acquired under 522 nm probing laser excitation at 100 µW power) shows several diffraction-limited spots surrounding the irradiated spots, attributed to the spatial gaussian profile of the processing laser. The emission spectrum as well as the background-subtracted second-order autocorrelation function $g^{(2)}(t)$ (reported in **Fig. 5b** for the individual spot circled in white in **Fig. 5a**) evidenced the occurrence of single photon emitters with the spectral signature of the MgV⁻ center. The identification of individual emitters enabled quantifying the PL intensity $I_s$ at the single center level, which was determined as the integral of the ZPL spectral peak intensity averaged over 5 different emitters. This value was used as a normalization factor to quantify the number $N_1$ of MgV⁻ emitters fabricated at each processed spot, under the assumption of a linear scaling between the number of emitters and the PL emission intensity.

Considering the number $N_i$ of Mg ions introduced by ion implantation in the confocal volume as the product between the implantation fluence F=1x10¹³ cm⁻² and the confocal spot size given by the diffraction limit (A=7x10⁻² µm²), the $N_1/N_i$ ratio value a direct assessment of the MgV⁻ formation efficiency for the process, defined as the number of optically-active emitters per implanted ion. The calculated formation efficiency is plotted in **Fig. 5c** for all the considered processing wavelengths and deposited energies. The formation efficiency for the MgV⁻ center is thus lower bound to a 0.11±0.02 % value for the 522 nm processing wavelengths, and to a 0.16±0.02% and 0.18±0.02% values at 445 nm and 405 nm, respectively.

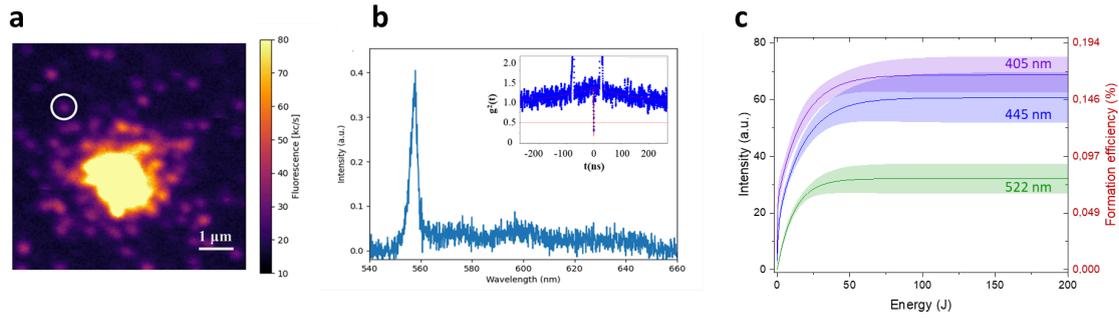

**Figure 5: a)** PL map of a confocal spot in Mg-implanted diamond processed under 445 nm laser wavelength (5 mW) for 10 h, surrounded by process-induced diffraction-limited PL emitting features, **b)** PL emission spectrum and second-order auto-correlation function acquired from the spot circled in white in Fig. 5a. **c)** MgV⁻ ZPL intensity and formation efficiency of the MgV– center as a function of the laser processing wavelength and deposited energy. The tinted areas describe the 95% confidence bands.

**Sn-related centers.** A similar characterization as the one reported for the MgV⁻ center was carried for Sn-related color centers. The processing parameters were chosen according to the same experimental details shown in **Table 1**. In **Figs. 6a-d** the arrays of spots processed at different laser processing wavelengths (namely: 522 nm, 445 nm, and 405 nm, respectively) are shown. Also in the case of the 595 nm Sn-related emission, the processing by means of a 594 nm laser source did not result in any activation of the color center. In **Figs. 6e-h** the corresponding PL intensity curves (extracted as the integral area underlying the 593 nm ZPL emission) are plotted versus the total energy deposited at each processing spot. The energy dependence shows a similar behavior with respect to what discussed in **Eqs. (1-5)** and **Figs. 3e-h** for the MgV⁻ center.

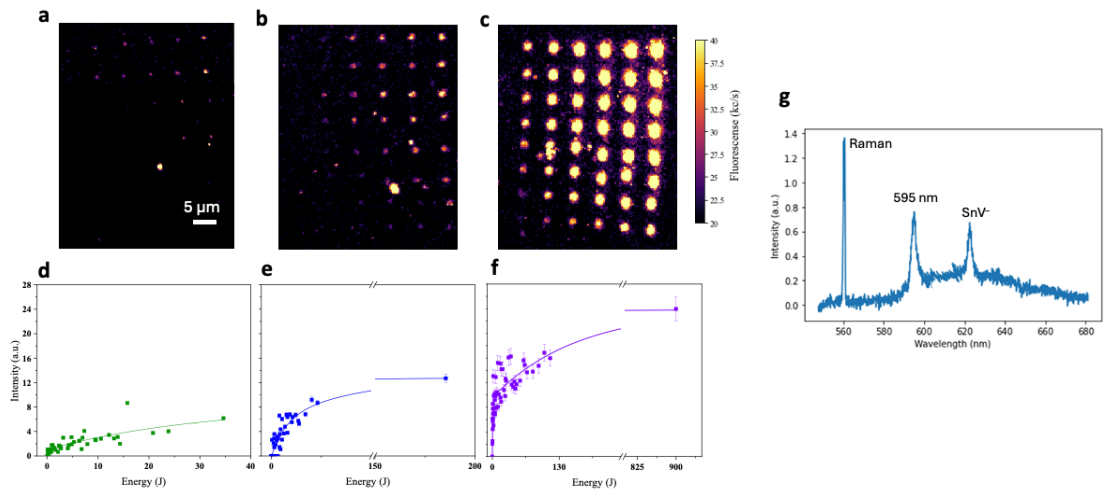

**Figure 6:** Photoactivation dependence on the laser processing parameters for the Sn-related 595 nm color center. The first row shows the confocal maps acquired on an array of spots undergone photoactivation via laser processing at **a)** 522 nm, **b)** 445 nm, **c)** 405 nm. The second row shows the integrated PL intensity (integral of the 595 nm ZPL spectral peak) collected spots as a function of the energy deposited on each spot under processing wavelength of **d)** 522 nm, **e)** 445 nm, **f)** 405 nm. **g)** spectrum acquired from a spot processed under 405 nm laser at (30 J delivered energy).

The curves were fitted according to the model described in **Eq. 5** and showed a similar decrease in the asymptotic PL emission intensity at increasing wavelengths, indicating a higher activation efficiency at higher processing photon energies. Differently from the results recorded for the MgV⁻ center, however, both the $I(E)$ curves corresponding to 522 nm and 445 nm laser processing could be suitably described with a single-exponential trend (i.e. $b=0$ in Eq. 5), which is compatible with a photoactivation process involving the ionization of a single dark charge state into the optically-active 595 nm emission. Conversely, the fitting of the $I(E)$ curve acquired related to the 405 nm processing wavelength required the introduction of a second exponential term according to **Eq. 5**. Also, the resulting exponential damping parameters (all fitting parameters are reported in **Table 3**) were in this case significantly modified with respect to the case of the other processing wavelengths, exhibiting a dominant contribution (i.e. ~60% of the emission intensity through the parameter $a$ in **Eq. 5**) from a process defined by $\alpha = 140 \pm 50$ J, paired with a second process described by $\beta = 0.8 \pm 0.2$ J. The need for a second exponential term could be together with an exhaustive description of the process would require a better understanding of the nature of the defect emitting at 595 nm. This emission was indeed reported in Sn-implanted diamond alongside with the SnV$^-$ ZPL [21], [34], but its properties are largely unexplored at the state of the art. Despite the lack of specific studies available on the photophysical properties of this defect, Iwasaki et al. [34] hypothesized its attribution to a bond-centered defect in a different structural configuration with respect to the SnV$^-$ center. This attribution is justified by the fact that the 595 nm optical center, differently from the SnV$^-$, anneals out at temperatures above 1400 °C [34]. This picture appears to be fully in line with the available structural analysis of the SnV center formation by emission channeling technique [33].

The interpretation of the 595 nm line as a different charge state of the SnV complex is further ruled out by the the occasional activation of the SnV$^-$ emission at 620 nm under the sole 405 nm laser processing (**Fig. 6g**). The observation of both PL emission lines could justify the significant variation in the $\alpha$ parameter under 405 nm processing due to the emergence of two competing ionization processes.

| Wavelength (nm) | $I_0$ (a.u.) | $a$ (a.u.) | $b$ (a.u.) | $\alpha$ (J) | $\beta$ (J) |
|---|---|---|---|---|---|
| 520 | 5.1±0.8 | - | - | 10±3 | - |
| 445 | 14.9±1.2 | - | - | 15±3 | |
| 405 | 24±3 | 14±3 | 8.7±0.7 | 140±50 | 0.8±0.2 |

**Table 3**. Fitting parameter of the wavelength-dependent $I(E)$ curves for the 595 nm Sn-related PL emission according to Eq. (5)

Similarly to what performed for the MgV$^-$ optical center, several individual Sn-related emitters were identified in the processed spots. An exemplary result is shown in **Fig. 7a**, where the PL map of the region surrounding a spot processed at 405 nm laser wavelength (900 J delivered energy) exhibits different isolated emitting spots which are diffraction-limited in size. The occurrence of single-photon emission from 595 nm centers (spectral signature shown in **Fig. 7b** for the emitter circled in white in **Fig. 7a**) was assessed by Hanbury-Brown & Twiss interferometry (**Fig. 7b**), showing a $g^{(2)}(t=0) = 0.214 \pm 0.032$ value under 80 μW probing laser excitation with a characteristic emission time of ~7.2 ns.

The formation efficiency was estimated with the same procedure adopted for the MgV$^-$ center for the considered laser processing wavelengths. The analysis yielded asymptotic efficiency values of 4.25±0.07 %, 2.11±0.04% and 1.23±0.02% for 405 nm, 445 nm and 522 nm processing wavelengths, respectively **Fig. 7c**. It is worth remarkig that the reported values, especially the one obtained under 405 nm irradiation, are satisfactorily compatible with what has been reported for other centers (SnV, MgV, GeV) under conventional ion irradiation and high temperature post-implantation annealing [43,44,45,46].

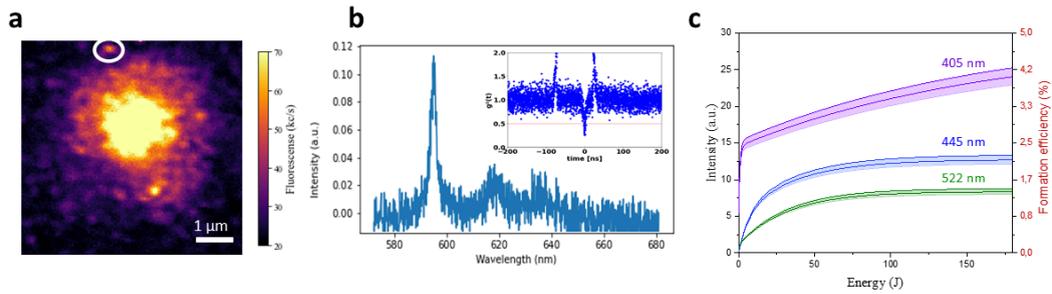

**Figure 7: a)** PL map of a confocal spot in Sn-implanted diamond processed under 445 nm laser wavelength (5 mW) for 10 h, surrounded by process-induced diffraction-limited PL emitting features, **b)** PL emission spectrum and second-order auto-correlation function acquired from the spot circled in white in Fig. 7a. **c)** 595 nm spectrum intensity and formation efficiency of the 595 nm center as a function of the laser processing wavelength and deposited energy. The tinted areas describe the 95% confidence bands.

**Discussion**

In this work the optical activation of the MgV$^-$ and of the 595 nm Sn-related optical centers were demonstrated following ion implantation and laser processing with visible laser source at optical powers in the 1-25 mW range. The PL emission from these color centers was obtained without the need for conventional thermal treatments [25,26,27,28] after ion implantation. The activation of the centers was correlated to the energy density delivered by the laser source on the processed spots, and was interpreted in terms of charge state conversion of the considered point defects. The occasional activation of the SnV$^-$ center in Sn-implanted diamond was observed under 405 nm processing.

These findings, which are complementary with respect to what recently reported for laser-induced SiV$^0$-SiV$^-$ photodynamics [47,48], confirm that a fraction of the Mg- and Sn-related defects is found in the structural bond-centered configuration right after ion implantation, as already observed in previous studies [19], [33]. Conversely, the formation efficiency estimated for all the considered defect classes (~1%) suggests that other factors may affect the optical activation of structurally stable defects, such as the effect on the Fermi level local environment in implanted, radiation-damaged diamond [49,50].

Although rather low, the non-negligible formation efficiency associated with these laser-based processes enables us to identify a strategy for the controlled activation of quantum-light emitters in diamond by loosening the stringent requirements on single-ion implantation [51,52]. Firstly, in this work it was shown that the ion implantation did not result in the appearance of background photons, since no detectable signal is observed right after ion implantation, and that the color centers activated by laser processing display a stable emission over months of probing time. Secondly, the activation was found to be spatially constrained to the focussing spot of the processing laser, and it was achieved by exploiting the same probing confocal microscope needed for the subsequent excitation and optical manipulation of the emitters. Thirdly, the formation of individual emitters was directly monitored by detecting discrete increases in the PL emission count rate detected by the confocal microscope itself. Thus, a fabrication protocol can be devised, in which an adequate ion fluence (e.g., ~100 ions per implanted spot, according to the formation efficiency) is combined with the real-time, *in situ* laser processing technique to achieve the activation of individual color centers at specific positions of a large area sample. Such a lithographic system, based on a confocal microscope, is simple and offers the advantage of spatial resolution with respect to conventional thermal annealing without requiring challenging focussing or collimation procedures for the impinging ion implantation beam.

Furthermore, the experimental apparatus is relatively inexpensive with respect to the exploitation of more elaborate laser sources (e.g., femtosecond lasers [53,54,55]) relying on irreversible local structural modification and annealing of the host material.

## Methods

**Sample preparation**. The measurements were conducted on two high-purity single-crystal IIa diamond substrates with <100> orientation, with nominal concentration of both substitutional B and N dopants the <5 ppb. Both samples were implanted using the multi-elemental negative ion implanter of the Solid State Physics laboratories of the University of Torino [56], [57]. A collimation system was used to define a ~ mm-sized implantation area. The ion fluence was estimated by the exposure time of the sample to the ion beam. The measurement of the ion current on a Faraday cup prior and after the implantation. The sample did not undergo subsequent annealing and was thus laser processed without previous surface chemical functionalization steps or thermal annealing.

**PL confocal microscopy**. The PL characterization of color centers at both the ensemble and single emitter level was performed using a custom single-photon sensitive confocal microscope. The setup was equipped with a 100x dry objective (0.9 NA). A set of fiber-coupled CW laser diodes (522 nm, 445 nm, 405 nm wavelengths) was exploited to deliver optical excitation. The photon collection was provided by coupling a pair of free-running, commercial Silicon single-photon avalanche diodes (Si-SPADs) (0.1 kcps dark count rate, ~45 ns deadtime, detection efficiency of ~50 % at 520 nm) to a fiber-fused multimode beamsplitter whose ø=50 μm core was exploited as the pinhole of the confocal microscope. The Si-SPADs pair fed a FPGA time-tagger board to enable the acquisition of second-orded autocorrelation measurements. A 550 nm long-pass dichroic mirror was employed to decouple the excitation photons from the PL signal. The spectral analysis of the PL emission was performed using a fiber-coupled SpectraPro HRS 300 spectrograph equipped with a PIXIS camera, (0.2 nm resolution, 330-800 nm spectral range). The PL emission from both samples was acquired by exploiting a 550 nm dichroic mirror along with a long-pass filter at 550 nm cutoff wavelength to filter out the laser excitation. For the sole spectral analysis carried on the laser-processed spots in $MgH^-$-implanted diamond, an additional 650 nm short-pass filter was employed to filter out the background PL originating from radiation-induced, photo-activated GR1 centers [40].

## Acknowledgements

This work was supported by the following projects: Project "Piemonte Quantum Enabling Technologies" (PiQuET), funded by the Piemonte Region within the "Infra-P" scheme (POR-FESR 2014-2020 program of the European Union); 'Training on LASer fabrication and ION implantation of DEFects as quantum emitters' (LasIonDef) project, funded by the European Research Council under the 'Marie Skłodowska-Curie Innovative Training Networks' program; "Departments of Excellence" (L. 232/2016), funded by the Italian Ministry of Education, University and Research (MIUR); "Ex post funding of research - 2021" of the University of Torino funded by the "Compagnia di San Paolo"; experiments ROUGE, QUISS funded by the 5th National Commission of the Italian National Institute for Nuclear Physics (INFN). We acknowledge financial support under the National Recovery and Resilience Plan (NRRP), Mission 4, Component 2, Investment 1.1, Call for tender No. 1409 published on 14.9.2022 by the Italian Ministry of University and Research (MUR), funded by the European Union – NextGenerationEU – Project  P2022KSTSR - Opto-mechanical effects in spin-defects for quantum technologies - CUP D53D23019370001. The project contributing to this work 23NRM04 NoQTeS has received funding from the European Partnership on Metrology, co-financed from the European Union's Horizon Europe Research and Innovation Programme and by the Participating States."

# REFERENCES

[1]  I. Aharonovich and E. Neu, "Diamond nanophotonics," *Adv Opt Mater*, vol. 2, no. 10, pp. 911–928, Oct. 2014, doi: 10.1002/adom.201400189.

[2]  M. Ruf, N. H. Wan, H. Choi, D. Englund, and R. Hanson, "Quantum networks based on color centers in diamond," *J Appl Phys*, vol. 130, no. 7, Feb. 2021, doi: 10.1063/5.0056534.

[3]  T. Schröder *et al.*, "Quantum nanophotonics in diamond [Invited]," *Journal of the Optical Society of America B*, vol. 33, no. 4, pp. B65–B83, 2016, doi: 10.1364/JOSAB.33.000B65.

[4]  C. Bradac, W. Gao, J. Forneris, M. E. Trusheim, and I. Aharonovich, "Quantum nanophotonics with group IV defects in diamond," *Nat Commun*, vol. 10, no. 1, p. 5625, Feb. 2019, doi: 10.1038/s41467-019-13332-w.

[5]  W. B. Gao, A. Imamoglu, H. Bernien, and R. Hanson, "Coherent manipulation, measurement and entanglement of individual solid-state spins using optical fields," *Nat Photonics*, vol. 9, no. 6, pp. 363–373, Feb. 2015, doi: 10.1038/nphoton.2015.58.

[6]  M. W. Doherty, N. B. Manson, P. Delaney, F. Jelezko, J. Wrachtrup, and L. C. L. L. Hollenberg, "The nitrogen-vacancy colour centre in diamond," *Phys Rep*, vol. 528, no. 1, pp. 1–45, 2013, doi: 10.1016/j.physrep.2013.02.001.

[7]  K. Tsurumoto, R. Kuroiwa, H. Kano, Y. Sekiguchi, and H. Kosaka, "Quantum teleportation-based state transfer of photon polarization into a carbon spin in diamond," *Commun Phys*, vol. 2, no. 1, p. 74, Jun. 2019, doi: 10.1038/s42005-019-0158-0.

[8]  F. Dolde *et al.*, "Room-temperature entanglement between single defect spins in diamond," *Nat. Phys.*, vol. 9, no. 3, pp. 139–143, 2013, doi: 10.1038/nphys2545.

[9]  R. Hanson, "Optimized quantum sensing with a single electron spin using real-time adaptive measurements," *Nat. Nanotechnol.*, no. November, 2015, doi: 10.1038/nnano.2015.261.

[10]  D. M. Toyli, C. D. Weis, G. D. Fuchs, T. Schenkel, and D. D. Awschalom, "Chip-Scale Nanofabrication of Single Spins and Spin Arrays in Diamond," *Nano Lett*, vol. 10, no. 8, pp. 3168–3172, Feb. 2010, doi: 10.1021/nl102066q.

[11]  A. M. Wojciechowski *et al.*, "Precision temperature sensing in the presence of magnetic field noise and vice-versa using nitrogen-vacancy centers in diamond," *Appl Phys Lett*, vol. 113, no. 1, p. 13502, Feb. 2018, doi: 10.1063/1.5026678.

[12]  E. H. Chen *et al.*, "High-sensitivity spin-based electrometry with an ensemble of nitrogen-vacancy centers in diamond," *Phys Rev A  (Coll Park)*, vol. 95, no. 5, p. 53417, Feb. 2017, doi: 10.1103/PhysRevA.95.053417.

[13]  D. D. Awschalom, R. Hanson, J. Wrachtrup, and B. B. Zhou, "Quantum technologies with optically interfaced solid-state spins," *Nat Photonics*, vol. 12, no. 9, pp. 516–527, Sep. 2018, doi: 10.1038/s41566-018-0232-2.

[14]  M. Alkahtani *et al.*, "Tin-vacancy in diamonds for luminescent thermometry," *Appl Phys Lett*, vol. 112, no. 24, p. 241902, Feb. 2018, doi: 10.1063/1.5037053.

[15]  A. E. Rugar *et al.*, "Quantum Photonic Interface for Tin-Vacancy Centers in Diamond," *Phys Rev X*, vol. 11, no. 3, p. 31021, Feb. 2021, doi: 10.1103/PhysRevX.11.031021.

[16]    E. I. Rosenthal *et al.*, "Microwave Spin Control of a Tin-Vacancy Qubit in Diamond," *Phys Rev X*, vol. 13, no. 3, p. 31022, Feb. 2023, doi: 10.1103/PhysRevX.13.031022.

[17]    B. Pingault, T. Muller, C. Hepp, E. Neu, C. Becher, and M. Atature, "Optical signatures of spin in silicon-vacancy centre in diamond," in *Optics InfoBase Conference Papers*, 2014. doi: 10.1364/cleo_qels.2014.fw1b.1.

[18]    M. Leifgen *et al.*, "Evaluation of nitrogen- and silicon-vacancy defect centres as single photon sources in quantum key distribution," *New J Phys*, vol. 16, no. 2, p. 23021, Feb. 2014, doi: 10.1088/1367-2630/16/2/023021.

[19]    E. Corte *et al.*, "Magnesium-Vacancy Optical Centers in Diamond," *ACS Photonics*, vol. 10, no. 1, pp. 101–110, Jan. 2023, doi: 10.1021/acsphotonics.2c01130.

[20]    G. Thiering and A. Gali, "Ab Initio Magneto-Optical Spectrum of Group-IV Vacancy Color Centers in Diamond," *Phys Rev X*, vol. 8, no. 2, p. 21063, Feb. 2018, doi: 10.1103/PhysRevX.8.021063.

[21]    S. D. D. Tchernij *et al.*, "Single-Photon-Emitting Optical Centers in Diamond Fabricated upon Sn Implantation," *ACS Photonics*, vol. 4, no. 10, pp. 2580–2586, Feb. 2017, doi: 10.1021/acsphotonics.7b00904.

[22]    J. Görlitz *et al.*, "Spectroscopic investigations of negatively charged tin-vacancy centres in diamond," *New J Phys*, vol. 22, no. 1, p. 13048, Feb. 2020, doi: 10.1088/1367-2630/ab6631.

[23]    J. Görlitz *et al.*, "Coherence of a charge stabilised tin-vacancy spin in diamond," *npj Quantum Inf*, vol. 8, no. 1, Dec. 2022, doi: 10.1038/s41534-022-00552-0.

[24]    A. Pershin, G. Barcza, Ö. Legeza, and A. Gali, "Highly tunable magneto-optical response from magnesium-vacancy color centers in diamond," *npj Quantum Inf*, vol. 7, no. 1, p. 99, Feb. 2021, doi: 10.1038/s41534-021-00439-6.

[25]    S. Pezzagna *et al.*, "Creation of colour centres in diamond by collimated ion-implantation through nano-channels in mica," Feb. 2011, *John Wiley & Sons, Ltd*. doi: 10.1002/pssa.201100455.

[26]    J. F. Prins, "Ion-implanted, shallow-energy, donor centres in diamond: the effect of negative electron affinity," *Nucl Instrum Methods Phys Res A*, vol. 514, no. 1–3, pp. 69–78, Feb. 2003, doi: 10.1016/j.nima.2003.08.085.

[27]    E. N. Hernández *et al.*, "Efficiency Optimization of Ge‑V Quantum Emitters in Single‑Crystal Diamond upon Ion Implantation and HPHT Annealing," *Adv Quantum Technol*, vol. 6, no. 8, Feb. 2023, doi: 10.1002/qute.202300010.

[28]    E. Nieto Hernandez *et al.*, "Efficient Fabrication of High‑Density Ensembles of Color Centers via Ion Implantation on a Hot Diamond Substrate," *Advanced Physics Research*, Jul. 2024, doi: 10.1002/apxr.202400067.

[29]    J. Van Donkelaar *et al.*, "Single atom devices by ion implantation," *J. Phys. Condens. Matter*, vol. 27, no. 15, 2015, doi: 10.1088/0953-8984/27/15/154204.

[30]    S. Becker, N. Raatz, St. Jankuhn, R. John, and J. Meijer, "Nitrogen implantation with a scanning electron microscope," *Sci. Rep.*, vol. 8, no. 1, p. 32, Feb. 2018, doi: 10.1038/s41598-017-18373-z.

[31]	B. Naydenov *et al.*, "Engineering single photon emitters by ion implantation in diamond," *Appl Phys Lett*, vol. 95, no. 18, p. 181109, Feb. 2009, doi: 10.1063/1.3257976.

[32]	T. Lühmann, R. John, R. Wunderlich, J. Meijer, and S. Pezzagna, "Coulomb-driven single defect engineering for scalable qubits and spin sensors in diamond," *Nat Commun*, vol. 10, no. 1, p. 4956, Feb. 2019, doi: 10.1038/s41467-019-12556-0.

[33]	U. Wahl *et al.*, "Direct Structural Identification and Quantification of the Split-Vacancy Configuration for Implanted Sn in Diamond," *Phys Rev Lett*, vol. 125, no. 4, p. 045301, Jul. 2020, doi: 10.1103/PhysRevLett.125.045301.

[34]	T. Iwasaki *et al.*, "Tin-Vacancy Quantum Emitters in Diamond," *Phys Rev Lett*, vol. 119, no. 25, p. 253601, Dec. 2017, doi: 10.1103/PhysRevLett.119.253601.

[35]	S. Sciortino *et al.*, "Micro-beam and pulsed laser beam techniques for the micro-fabrication of diamond surface and bulk structures," *Nucl Instrum Methods Phys Res B*, vol. 348, pp. 191–198, Apr. 2015, doi: 10.1016/j.nimb.2014.11.061.

[36]	Y. Gao, J.-M. Lai, Z.-Y. Li, P.-H. Tan, C.-X. Shan, and J. Zhang, "Local laser heating effects in diamond probed by photoluminescence of SiV− centers at low temperatures," *Appl Phys Lett*, vol. 124, no. 9, Feb. 2024, doi: 10.1063/5.0184331.

[37]	C. Pederson *et al.*, "Optical tuning of the diamond Fermi level measured by correlated scanning probe microscopy and quantum defect spectroscopy," *Phys Rev Mater*, vol. 8, no. 3, p. 036201, Mar. 2024, doi: 10.1103/PhysRevMaterials.8.036201.

[38]	R. Kumar *et al.*, "Stability of Near-Surface Nitrogen Vacancy Centers Using Dielectric Surface Passivation," *ACS Photonics*, vol. 11, no. 3, pp. 1244–1251, Mar. 2024, doi: 10.1021/acsphotonics.3c01773.

[39]	D. J. Twitchen, D. C. Hunt, V. Smart, M. E. Newton, and J. M. Baker, "Correlation between ND1 optical absorption and the concentration of negative vacancies determined by electron paramagnetic resonance (EPR)," *Diam Relat Mater*, vol. 8, no. 8–9, pp. 1572–1575, Aug. 1999, doi: 10.1016/S0925-9635(99)00038-2.

[40]	A. Mainwood and A. M. Stoneham, "Stability of electronic states of the vacancy in diamond," *Journal of Physics: Condensed Matter*, vol. 9, no. 11, pp. 2453–2464, Mar. 1997, doi: 10.1088/0953-8984/9/11/013.

[41]	N. Aslam, G. Waldherr, P. Neumann, F. Jelezko, and J. Wrachtrup, "Photo-induced ionization dynamics of the nitrogen vacancy defect in diamond investigated by single-shot charge state detection," *New J Phys*, vol. 15, no. 1, p. 013064, Jan. 2013, doi: 10.1088/1367-2630/15/1/013064.

[42]	N. B. B. Manson and J. P. P. Harrison, "Photo-ionization of the nitrogen-vacancy center in diamond," *Diam Relat Mater*, vol. 14, no. 10, pp. 1705–1710, Feb. 2005, doi: 10.1016/j.diamond.2005.06.027.

[43]	U. Wahl *et al.*, "Structural formation yield of GeV centers from implanted Ge in diamond," *Materials for Quantum Technology*, vol. 4, no. 2, p. 025101, Jun. 2024, doi: 10.1088/2633-4356/ad4b8d.

[44]	A. E. Rugar, C. Dory, S. Sun, and J. Vučković, "Characterization of optical and spin properties of single tin-vacancy centers in diamond nanopillars," *Phys Rev B*, vol. 99, no. 20, p. 205417, Feb. 2019, doi: 10.1103/PhysRevB.99.205417.

[45]	Y. Zhou *et al.*, "Direct writing of single germanium vacancy center arrays in diamond," *New J. Phys.*, vol. 20, no. 12, p. 125004, Feb. 2018, doi: 10.1088/1367-2630/aaf2ac.

[46]	T. Lühmann, J. Meijer, and S. Pezzagna, "Charge‑Assisted Engineering of Color Centers in Diamond," *physica status solidi (a)*, vol. 218, no. 5, Feb. 2021, doi: 10.1002/pssa.202000614.

[47]	Z.-H. Zhang *et al.*, "Neutral Silicon Vacancy Centers in Undoped Diamond via Surface Control," *Phys Rev Lett*, vol. 130, no. 16, p. 166902, Apr. 2023, doi: 10.1103/PhysRevLett.130.166902.

[48]	G. Garcia‑Arellano, G. I. López‑Morales, N. B. Manson, J. Flick, A. A. Wood, and C. A. Meriles, "Photo‑Induced Charge State Dynamics of the Neutral and Negatively Charged Silicon Vacancy Centers in Room‑Temperature Diamond," *Advanced Science*, vol. 11, no. 22, Jun. 2024, doi: 10.1002/advs.202308814.

[49]	A. T. Collins, "The Fermi level in diamond," *Journal of Physics: Condensed Matter*, vol. 14, no. 14, pp. 3743–3750, Feb. 2002, doi: 10.1088/0953-8984/14/14/307.

[50]	Y. Mita, "Change of absorption spectra in type-I *b* diamond with heavy neutron irradiation," *Phys Rev B*, vol. 53, no. 17, pp. 11360–11364, May 1996, doi: 10.1103/PhysRevB.53.11360.

[51]	D. N. Jamieson *et al.*, "Controlled shallow single-ion implantation in silicon using an active substrate for sub-20-keV ions," *Appl Phys Lett*, vol. 86, no. 20, p. 202101, Feb. 2005, doi: 10.1063/1.1925320.

[52]	T. Matsukawa, T. Fukai, S. Suzuki, K. Hara, M. Koh, and I. Ohdomari, "Development of single-ion implantation — controllability of implanted ion number," *Appl Surf Sci*, vol. 117–118, pp. 677–683, Feb. 1997, doi: 10.1016/S0169-4332(97)80163-8.

[53]	H. Quard *et al.*, "Femtosecond-laser-induced creation of G and W color centers in silicon-on-insulator substrates," *Phys Rev Appl*, vol. 21, no. 4, p. 044014, Apr. 2024, doi: 10.1103/PhysRevApplied.21.044014.

[54]	A. L. Robinson, "Femtosecond Laser Annealing of Silicon," *Science (1979)*, vol. 226, no. 4672, pp. 329–330, Feb. 1984, doi: 10.1126/science.226.4672.329.

[55]	J. Engel *et al.*, "Combining femtosecond laser annealing and shallow ion implantation for local color center creation in diamond," *Appl Phys Lett*, vol. 122, no. 23, Jun. 2023, doi: 10.1063/5.0143922.

[56]	E. Nieto Hernández *et al.*, "Fabrication of quantum emitters in aluminum nitride by Al-ion implantation and thermal annealing," *Appl Phys Lett*, vol. 124, no. 12, Mar. 2024, doi: 10.1063/5.0185534.

[57]	G. Andrini *et al.*, "Activation of telecom emitters in silicon upon ion implantation and ns pulsed laser annealing," *Commun Mater*, vol. 5, no. 1, p. 47, Apr. 2024, doi: 10.1038/s43246-024-00486-4.